\documentclass{XrU2005}

\def\ecs{ergs cm$^{-2}$ s$^{-1} \ $}
\def\es{ergs s$^{-1}\ $}

\def\lae{\mathrel{<\kern-1.0em\lower0.9ex\hbox{$\sim$}}}
\def\gae{\mathrel{>\kern-1.0em\lower0.9ex\hbox{$\sim$}}}

\title{The Bright Side of the hard X-ray Sky: The XMM-Newton Bright Survey}
\author[1,3]{R. Della Ceca}
\author[1,3]{ A. Caccianiga}
\author[1,3]{P. Severgnini}
\author[1,3]{T. Maccacaro}
\author[1,3]{ F. Cocchia}
\author[2,3]{ V. Braito}
\affil[1]{INAF-Osservatorio Astronomico di Brera, via Brera 28, I-20121 Milan, Italy}
\affil[2]{Johns Hopkins University, 3400 N. Charles Street, Baltimore, MD 21218,USA}
\affil[3]{{\bf on behalf of the XMM-Newton SSC collaboration}}

\begin{document}


\maketitle

\begin{abstract}
We discuss here the main goals and some interesting results 
of the ``XMM-Newton Bright Serendipitous Survey", 
a research program conducted by the XMM-Newton Survey Science Center
\footnote{The MM-Newton Survey Science Center (SSC,  see
http://xmmssc-www.star.le.ac.uk.) is an international collaboration, involving a
consortium of 10 institutions, appointed by ESA to help the SOC in developing
the software analysis system, to pipeline  process all the XMM-Newton data, and 
to exploit the XMM serendipitous detections. The OABrera is one of the
Consortium Institutes.}
in two 
complementary energy bands (0.5--4.5 keV and 4.5--7.5 keV) in the bright 
(above $\sim 7 \times 10^{-14}$ erg cm$^{-2}$ s$^{-1}$) flux regime.
 The very high identification rate (96\%) for the X-ray source sample selected 
 in 
the 4.5--7.5 keV band is used here to have, in this energy range, 
 an ``unbiased" view  of the 
extragalactic hard X-ray sky at bright fluxes. 
\end{abstract}

\section{Introduction}

Deep {\it Chandra} and {\it XMM--Newton}  observations  have resolved $\sim
80$\%   and $\sim 60$\% of the 0.5--5 keV and 5--10 keV cosmic X-ray background
(CXB) into discrete sources down to   $f_{x}\sim 3\times$10$^{-16}$ erg
cm$^{-2}$ s$^{-1}$  and $f_{x}\sim 3\times$10$^{-15}$ erg cm$^{-2}$ s$^{-1}$,
respectively (\citealt{worsley2005} and reference therein).

The X--ray data  (stacked spectra and hardness ratios) of these faint samples
are consistent with AGN  being the dominant contributors to the CXB 
(\citealt{BH2005} and references therein; G. Hasinger and Y. Ueda
contributions to this meeting) and, as  inferred by the X--ray colors, a
significant fraction  of these sources have hard, presumably obscured, X--ray
spectra,  in agreement with the predictions of CXB synthesis models 
(\citealt{setti1989, comastri2001, gilli2001, ueda2003, treister05}).

On the other hand, the majority of the sources found in  medium to deep
fields  are too faint to provide good X--ray spectral information. Furthermore,
the extremely faint magnitudes  of a large number of their optical counterparts
make the spectroscopic identifications very difficult, or even impossible, with
the present day  ground--based optical telescopes. Thus, notwithstanding the
remarkable results obtained by reaching very faint X--ray fluxes, the
broad--band physical  properties (e.g. the relationship between optical
absorption and X-ray obscuration and the reason why AGN with similar X-ray
properties have completely different optical  appearance) are not yet completely
understood.  In the medium flux regime  a step forward toward the solution of
some of these problems  has been undertaken  by \citet{mainieri2002},
\citet{piconcelli2003}, \citet{perola2004} and \citet{mateos2005}.

With the aim of complementing the results obtained by medium to deep  X-ray surveys, we
have  built the ``The XMM-Newton Bright Serendipitous Source Sample" 
(\citealt{dellaceca2004}).  We describe below the
main characteristics of this sample and discuss some  of the results obtained so far.
The contribution of this project to the solution of some critical open (and ``hot")
questions like the relationship between optical absorption and X-ray obscuration  and
the physical nature of the ``X-ray bright optically normal galaxies" have been already
discussed in  \citet{caccianiga2004} and \citet{severgnini2003},  respectively. 
We stress that
many of these issues are investigated with difficulties when 
using the fainter X-ray samples
because of the typical poor counts statistics available for each source. 
Here we consider the cosmological model with 
($H_o$,$\Omega_M$,$\Omega_{\lambda}$)=(65,0.3,0.7).

\section{The XMM-Newton Bright Serendipitous Source Sample}

The XMM Bright Serendipitous Source Survey,  a project led by the
{\it Osservatorio Astronomico di Brera}, consists of two flux-limited samples: the
XMM BSS and the XMM HBSS sample having a flux limit of $\sim 7 \times 10^{-14}$
erg cm$^{-2}$ s$^{-1}$ in the 0.5--4.5 keV and 4.5--7.5 keV energy bands,
respectively.  This approach was determined by the need of studying the
composition of the source population as a function of the selection band
and in order to reduce the strong bias against absorbed sources.
which occurs when
selecting in soft X-rays.  

Two-hundred and thirty-seven  suitable XMM fields (211 for the HBSS) at
$|b|>20$ deg were analyzed and
a sample of 400 sources was selected (see Table 1 for details and 
\citealt{dellaceca2004}). 
\begin{table}
\begin{center}
\caption{The current optical breakdown of the BSS and HBSS
samples}
\begin{tabular}{lrr}
\hline
\hline
                     &           BSS         & HBSS         \\
                     &                       &              \\
                     &                       &              \\
\hline
Objects$^{1}$        &           389         &          67  \\
Area Covered (deg$^2$) &          28.10      &          25.17 \\
\\
Identified           &           308         &          64  \\
Identification rate  &           79\%        &          96\% \\
                     &                       &              \\
AGN-1                &           201         &          38  \\
AGN-2                &            29         &          20  \\
Galaxies$^{2}$       &             8         &           1  \\
Clusters of Galaxies $^{3}$ &      6         &           1  \\
BL Lacs              &             6         &           2   \\
Stars$^{4}$          &            58         &           2   \\
\hline
\hline
\end{tabular}
\end{center}
$^{1}$  Fifty-six sources are in common between the BSS and HBSS
samples;
$^{2}$  we stress that some of the sources classified as ``Optically Normal
Galaxy" could indeed host an optically elusive AGN (\citealt{severgnini2003});
$^{3}$  the source detection algorithm is optimized for point-like 
objects, so the sample of clusters of galaxies is not statistically complete nor 
representative of the cluster population;
$^{4}$  all but one 
of the sources classified as stars are coronal emitters. The 
stellar content of the XMM-BSS has been presented to this meeting 
by J. Lopez-Santiago.
\end{table}
It is worth noting that among the ongoing surveys performed with {\it Chandra}
and XMM-Newton the XMM BSS survey is currently  covering the 
largest area; furthermore,  unlike deep pencil beam surveys, the XMM BSS
is unbiased by problems connected to the cosmic variance.

The majority of the BSS X-ray sources have enough
statistics (hundreds to  thousands counts when the data from the three EPIC
detectors are considered) to allow X-ray studies in terms of energy
distributions,  absorption properties, source extent and flux variability.
Moreover  the optical counterpart of $\sim 90$\%  of the X-ray sources has a
magnitude brighter than the POSS II limit (R $\sim$ 21mag), thus allowing
spectroscopic identification at a 2-4 meter class telescope;  this fact, combined with
the XMM positional accuracy for bright sources (90\% error circle of 
4$^{\prime\prime}$) implies that, in almost all cases, only one
object  needs to be observed to secure the optical identification.

Up to now 318 X-ray sources have been spectroscopically identified 
(220 sources from  our own observations and the remaining 
from the literature) leading to a 79\% and
96\% identification rate for the BSS and HBSS samples respectively (see Table 1
for a  summary). 
The optical selection criteria for the spectroscopic classification 
have 
been discussed in \citet{caccianiga2004} and \citet{dellaceca2004}; a 
full description of the
optical properties of the sources will be presented 
in Caccianiga et al., in preparation. 

\begin{figure}
\centering
\includegraphics[width=0.95\linewidth]{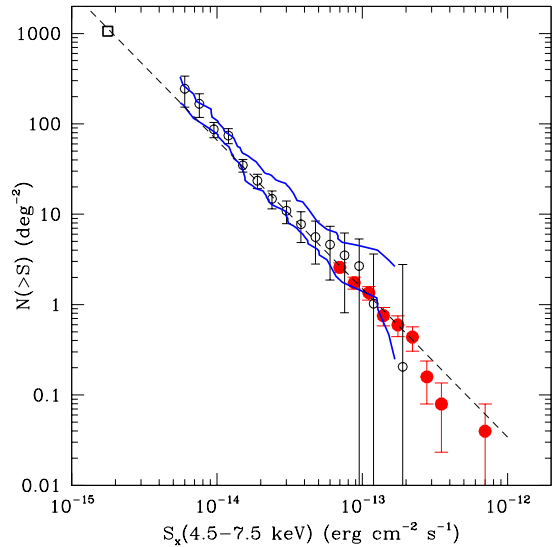}
%\vspace{-2cm}
\caption
{The extragalactic number-flux relationship in the 4.5--7.5 keV 
energy band obtained using the HBSS sample (binned
representation: filled circles). 
We have also reported for comparison other 
XMM-Newton based Log(N$>$S)$-$LogS, obtained in  
a similar energy range, converted to the 4.5--7.5 keV
energy range. In particular we have reported 
the HELLAS2XMM 5--10 keV Log(N$>$S)$-$LogS (area inside the
thick solid lines, \citealt{baldi2002}) and  
the AXIS 4.5--7.5 keV  Log(N$>$S)$-$LogS (open circles, 
Carrera \& Barcons 2005, private communication). The open square at $S
\simeq 2\times 10^{-15}$ \ecs represents the extragalactic surface
density  in the Lockman hole field according to \citet{hasinger2001}.\label{fig:fig1}
The dashed line represents the  best fit Log(N$>$S)$-$LogS obtained using 
the HBSS source sample (\citealt{dellaceca2004});  
for comparison the power-law best fit model has been extrapolated 
down to $\simeq 10^{-15}$ \ecs.
}
\end{figure}

\section{The HBSS sample}

The HBSS sample is now almost completely identified ($96\%$ of spectroscopic 
identifications with only three sources currently unidentified), 
therefore it can be used 
to have an 创unbiased创 view of the high  galactic 
($|b|>20$ deg) 4.5--7.5 keV  sky 
in the bright flux regime. 
First results, based on a complete subsample of 28 objects completely 
identified, have been already discussed in \citet{caccianiga2004}.
We stress that an identification rate around 80\% (as e.g. in the BSS sample or 
in other samples appeared in literature) could be 
not enough to have this 创unbiased创 view. Indeed, interesting classes of 
X-ray emitting sources (e.g. the narrow line AGN population), which
could represent an important minority of the source population, 
could be more difficult to identify.
We concentrate here the discussion on three main results concerning 
the HBSS sample: 
the 4.5--7.5 keV extragalactic number-flux relationship, the position 
of the AGN sample in the $L_x$-$N_H$ plane, and the correlation between 
X-ray colours and intrinsic $N_H$.

\subsection{The 4.5--7.5 keV extragalactic number-flux relationship}

In Fig.~\ref{fig:fig1} we compare the extragalactic number-flux relationship 
in the 4.5--7.5 keV energy band obtained using the HBSS sample (binned
representation: filled circles) with other 
XMM-Newton based Log(N$>$S)$-$LogS, obtained in  
a similar energy domain: the agreement is 
very good  making us confident about the 
reliability of the data selection and analysis. 
Furthermore, since we are covering the largest area in the bright 
flux regime we are able to set the 
most stringent constraints on the extragalactic source surface densities 
above $S\simeq 7\times 10^{-14}$ \ecs.
In Fig.~\ref{fig:fig1} we have also reported the 
best fit power-law Log(N$>$S)$-$LogS obtained using 
the HBSS source sample (\citealt{dellaceca2004}) extrapolating 
it down to $\simeq 10^{-15}$ \ecs; it is clear 
that we have no compelling evidence of 
a flattening of the 4.5--7.5 keV energy extragalactic 
number-flux relationship down to the 
flux limits sampled so far.

\subsection{The $L_x-N_H$ plane}

\begin{figure*}
\centering
\includegraphics[width=0.95\linewidth]{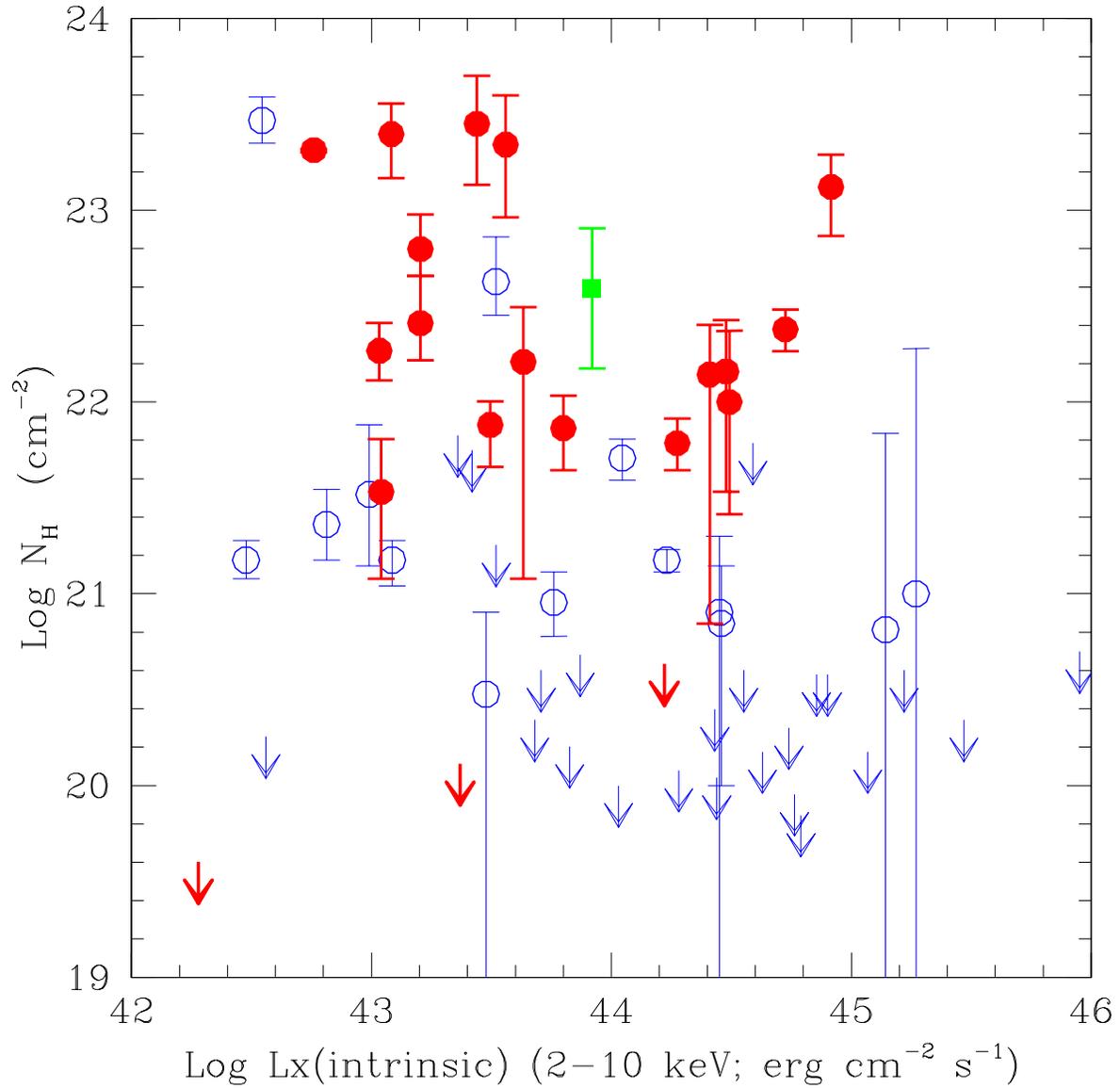}
%\vspace{4cm}
\caption
{Intrinsic 2-10 keV luminosity versus intrinsic absorption column densities (both quantities derived
from a complete spectral analysis) for the sources belonging to the HBS sample.
We have used different symbols to mark the different optical classification
of the objects.
The arrows indicate the 90\% upper limit on $N_H$.
Optically Type 1 AGN: open circles and thin arrows.
Optically Type 2 AGN: filled circles and thick arrows.
Optically normal galaxies: filled squares.\label{fig:fig2}}
\end{figure*}

In Fig.~\ref{fig:fig2} we show the intrinsic $L_{(2-10 {\it keV})}$ versus
 intrinsic $N_H$  
for the sources belonging to the HBSS sample.
Both the  intrinsic 2-10 keV luminosity (e.g. de-absorbed from the measured 
$N_H$ at the source redshift) and the intrinsic  $N_H$ 
have been derived from a complete X-ray spectral analysis using, 
when possible, data from all the three instruments on-board XMM-Newton 
(MOS1, MOS2, pn).
 
As already discussed in \citet{caccianiga2004} some 创hot创 questions  
could be investigated from a close inspection of Fig.~\ref{fig:fig2}, {\bf especially 
considering the fact that we are dealing with a well defined and complete 
sample which has been almost completely identified.} \\ \noindent
The main results can be summarized in these points:

$\bullet$ We do not find a strong evidence of a large population  of absorbed
($N_H>10^{22}$ cm$^{-2}$) optically Type 1  AGN. We only have two objects
belonging to this category,   i.e. about $3\%$ of the total extragalactic
population  and $5\%$ of the Type 1 AGN population shining in the 4.5--7.5 keV
sky  at the sampled fluxes. For one of these two objects the X-ray spectra
could also be  described by a typical Type 1 AGN power-law model  with a large
relativistic iron line;   a deeper investigation (using all the data in the  XMM
archive) is in progress to asses the presence  of intrinsic absorption. The
other absorbed object is  a nearby (z=0.019) and well known Narrow Line  Seyfert
1 galaxy (MKN 1239) studied in X-ray  also 
by \citet{grupe2004} using pointed XMM observations;

$\bullet$ Among optically Type 2 AGN about 65\% are characterized by an  
$N_H > 10^{22}$ cm$^{-2}$, 20\%  by an  $N_H$ between $10^{21.5}$ and
$10^{22}$ cm$^{-2}$ and  15\% are apparently unabsorbed ($N_H<10^{21.5}$
cm$^{-2}$; the three objects  marked with thick arrows). 
Two unabsorbed
sources have an intrinsic luminosity of  $\sim 2\times 10^{42}$ \es and  $\sim
2\times 10^{43}$ \es and are both classified as Seyfert 1.9.  
The latter unabsorbed Type 2 AGN is
apparently a high luminosity ($L_X \sim 2\times 10^{44}$ \es) narrow line AGN
although, unfortunately, the  $H_{\alpha}$ line is not sampled  at the moment.
Similar unabsorbed Type 2 QSOs have been discussed in \citet{wolter2005}
(see also the A. Wolter contribution to this meeting);

$\bullet$ As already found by other surveys we note a strong
deficiency of the number of high luminosity ($L_x > 10^{44}$ erg s$^{-1}$)
absorbed  ($N_H>10^{22}$ cm$^{-2}$) AGN, the so called absorbed Type 2 
QSOs predicted to be a consistent number 
by the synthesis models of the CXB. The
HBSS sample lists 5 Type 2 QSOs with 4 of them  having an absorbing column
density in a narrow range between  $10^{22}$ cm$^{-2}$  and $\sim 3\times
10^{22}$ cm$^{-2}$;

$\bullet$ At the fluxes covered by the HBSS survey 
the 4.5--7.5 keV selection is an
efficient way to sample AGN with absorbing column densities up  to $N_H\simeq
10^{24}$ cm$^{-2}$. Furthermore  we have no strong evidence of 
$N_H>10^{24}$ cm$^{-2}$ (e.g.  the presence of a strong iron line) for the three
optically  Type 2 AGN that are unabsorbed in the X-ray regime. Therefore, 
unless some
of our absorbed  objects are characterized by a dual absorber model (with an
absorbed component not visible below  10 keV) we do not find Compton thick AGN.
A new, unexpected and interesting result is that  very few
Compton thick AGN seem also 
to emerge from the Swift/BAT and INTEGRAL surveys of the 
hard (20-100 kev) sky above a flux limit of $10^{-11}$ \ecs (see V. Beckmann and 
R. Mushotzky contributions to this meeting). 
We are now
evaluating if the results from the HBSS can be used to constraint  the density
of the elusive Compton thick AGN at fainter fluxes;

$\bullet$ We only have one object that is optically classified 
as normal galaxy, but the $H{\alpha}$ line is not sampled at the moment. 
Its absorbing column density and luminosity are highly 
indicative of the presence of an absorbed AGN. A deeper 
investigation is in progress.

\subsection{X-ray colors versus absorbing column densities}

One of the responsibilities of the XMM-Newton SSC is the production of
the XMM-Newton Source Catalogue. This catalogue will provide a
rich and unique resource for generating well-defined samples for
specific studies, considering the fact that X-ray selection is a highly
efficient way of selecting AGN, clusters of galaxies and active stars. 

Having completed the X-ray spectral analysis of the sources in the HBSS 
sample we can investigate the (expected) correlation 
between the hardness ratios and the absorbing column densities for the 
AGN population.  
We have used here the hardness ratios HR2 as defined from the XMM-Newton
pipeline processing :
~~~~~~\\
 \[
HR2={C(2-4.5\, {\rm keV})-C(0.5-2\, {\rm keV})\over C(2-4.5\, {\rm
keV})+C(0.5-2\, {\rm keV})}
\]
~~~~~~\\
where C(0.5$-$2\, {\rm keV}), C(2$-$4.5\, {\rm keV}) 
 are the ``PSF and vignetting corrected"
count rates in the 0.5$-$2 and  2$-$4.5  keV energy bands, respectively.
We stress that  a ``Hardness Ratio" is often the only X-ray spectral 
information available for the faintest
sources in the XMM-Newton catalogue, and thus, a ``calibration" in the
parameter space is needed to select ``clean" and well-defined samples.
With the data we have accumulated so far we can work out this ``calibration" 
for e.g. absorbed AGN.

\begin{figure*}
\centering
\includegraphics[width=0.95\linewidth]{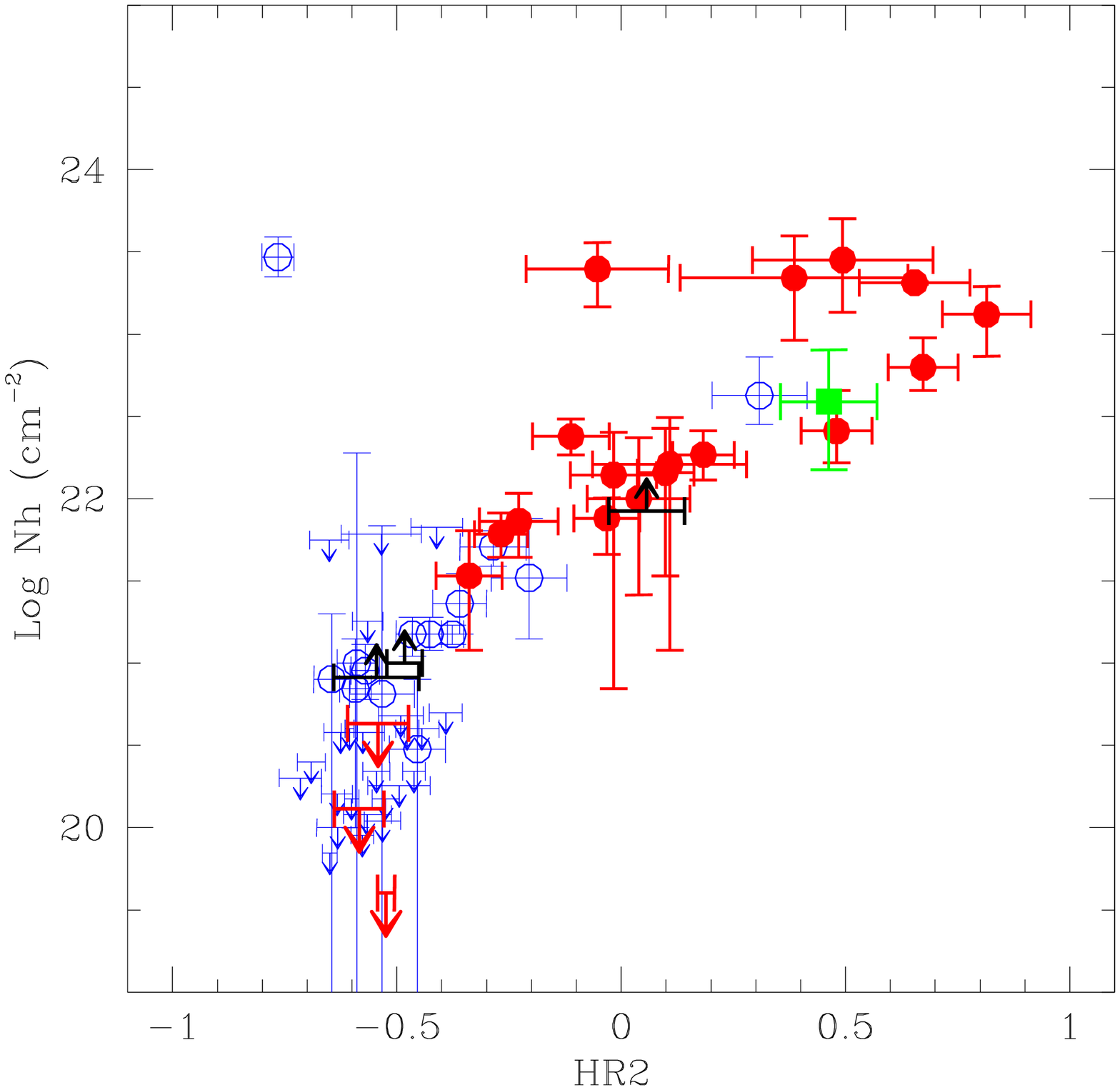}
%\vspace{-2cm}
\caption
{Intrinsic absorbing column densities versus their observed 
HR2 X-ray colours 
for the sources in the HBSS sample.
The arrows indicated the 90\% upper limit on $N_H$.
Symbols are as follows:
optically Type 1 AGN -- open circles and downward thin arrows;
optically Type 2 AGN -- filled circles and downward thick arrows;
optically normal galaxies -- filled squares;
unidentified objects -- upward thick arrows.
\label{fig:fig3}}
\end{figure*}

In Fig.~\ref{fig:fig3} we show the expected correlation  between the hardness ratio HR2,
the absorbing column densities and the optical spectral  properties of the
selected AGN. \\
\noindent
A few considerations are needed:

$\bullet$ The 4.5--7.5 keV selection, at the flux limits investigated here, seem
to be extremely efficient in selecting sources that are described 
at the ``first" order by an absorbed power-law model. 
This emerge from the consideration that there are very few objects
which  significantly deviate from the correlation between the hardness ratio HR2
and the $N_H$. A close inspection of the X-ray spectra of these few sources 
show
best fit spectral models more  complex than a simple absorbed power-law.  We have
some indications that  the situation could be a bit more complicated for the
sample of sources selected in the 0.5--4.5 keV energy band;

$\bullet$ We strengthen the results first discussed in \citet{caccianiga2004}
and \citet{dellaceca2004}.  First, all but one source characterized
by an intrinsic $N_H > 10^{21.5}$ cm$^{-2}$,   
have HR2$>-0.35$. Second, all
but 3 of the objects optically classified as  Type 2 AGN have HR2$>-0.35$.
Third, only three objects classified  as optically Type 1 AGN have HR2$>-0.35$
and only one of them have (but see the discussion above)  $N_H> 10^{22}$
cm$^{-2}$.

The results shown in Fig.~\ref{fig:fig3} allow us to design a 
very simple process to
pinpoint absorbed AGN {\em with  very  high efficiency}
using XMM-Newton data:
selection in the 4.5--7.5 keV energy band combined  with the condition that the
selected sources should have HR2$>-0.35$.  The fact that we are dealing with a
well defined and complete  sample which has been almost completely
identified allows us to  make a step forward: from a simple selection of
sources to a sample with  well defined statistical properties. 
In fact,  using the current HBSS sample as an ``unbiased" view of the X-ray sky 
in the  4.5--7.5
keV energy band, we gain information on which kind of sources   
we are leaving out by imposing the constraint of HR2$>-0.35$.    
The incoming 2XMM catalog will allow us to build up a sample of
about 100  absorbed AGN (from the analysis of about 1000 useful XMM fields) 
having fluxes above $\sim 10^{-13}$ \ecs.  Extrapolating the results obtained so far
on the HBSS (Della Ceca et al. in preparation) these sources should
allow us to study the cosmological evolution properties (XLF and evolution) 
of the absorbed AGN population in the luminosity range from 
$10^{42}$ to $10^{45}$ \es and up to z=1, 
i.e. a useful redshift and luminosity range 
to investigate their connection with galaxy 
evolution and star formation in the Universe
(\citealt{ballantyne2005}).

\section*{Acknowledgments}

We acknowledge partial financial support by  
ASI (grants: I/R/062/02 and I/R/071/02), by  MURST
(Cofin-03-02-23) and by INAF. We thank the TNG and the ESO NTT 
Time Allocation Committee for a generous and
continuous  allocation of observing time.

\end{document}